\documentclass[10pt,conference]{IEEEtran}
\usepackage{cite,graphicx,psfrag,amsmath,amssymb}
\newtheorem{example}{Example}

\begin{document}
\title{Efficient Implementation of the Generalized Tunstall Code Generation Algorithm}
\bibliographystyle{IEEEtran}
\author{\authorblockN{Michael~B.~Baer}
\authorblockA{VMware, Inc.\\
71 Stevenson St., 13\textsuperscript{th} Floor\\
San Francisco, CA  94105-0901, USA\\
Email: mbaer\hspace{1sp}@\hspace{1sp}vmware.com}}
\maketitle

\begin{abstract}
A method is presented for constructing a Tunstall code that is linear time in the number of output items.  This is an improvement on the state of the art for non-Bernoulli sources, including Markov sources, which require a (suboptimal) generalization of Tunstall's algorithm proposed by Savari and analytically examined by Tabus and Rissanen.  In general, if $n$ is the total number of output leaves across all Tunstall trees, $s$ is the number of trees (states), and $D$ is the number of leaves of each internal node, then this method takes $O((1+(\log s)/D) n)$ time and $O(n)$ space.
\end{abstract}

\section{Introduction}
\label{intro}

Although not as well known as Huffman's optimal fixed-to-variable-length coding method, the optimal variable-to-fixed-length coding technique proposed by Tunstall\cite{Tuns} offers an alternative method of block coding.  In this case, the input blocks are variable in size and the output size is fixed, rather than vice versa.  Consider a variable-to-fixed-length code for an independent, identically distributed (\textit{i.i.d.}) sequence of random variables $\{X_k\}$, where, without loss of generality, $\mbox{Pr}[X_k = i] = p_i$ for $i \in {\mathcal X} \triangleq \{0, 1, \ldots, D-1\}$.  The outputs are $m$-ary blocks of size $n=m^\nu$ for integers~$m$ --- most commonly $2$ --- and $\nu$, so that the output alphabet can be considered with an index $j \in {\mathcal Y} \triangleq \{0, 1, \ldots, n-1\}$. 

\begin{figure}[ht]
\centering
\psfrag{0}{\small $0_3$:}
\psfrag{10}{\small $10_3$:}
\psfrag{11}{\small $11_3$:}
\psfrag{12}{\small $12_3$:}
\psfrag{20}{\small $20_3$:}
\psfrag{21}{\small $21_3$:}
\psfrag{22}{\small $22_3$:}
\psfrag{sin 1}{\tiny $000_2$}
\psfrag{sin 2}{\tiny $001_2$}
\psfrag{sin 3}{\tiny $010_2$}
\psfrag{sin 4}{\tiny $011_2$}
\psfrag{sin 5}{\tiny $100_2$}
\psfrag{sin 6}{\tiny $101_2$}
\psfrag{sin 7}{\tiny $110_2$}
\includegraphics{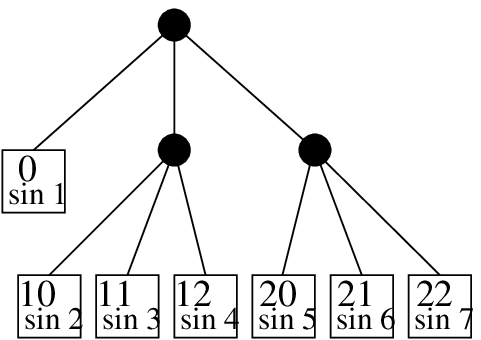}
\caption{Ternary-input $3$-bit-output Tunstall tree}
\label{parsetree}
\end{figure}

Codewords have the form ${\mathcal X}^*$, so that the code is a $D$-ary prefix code, suggesting the use of a coding tree.  Unlike the Huffman tree, in this case the inputs are the codewords and the outputs the indices, rather than the reverse; it thus parses the input and is known as a \textit{parsing tree}.  For example, consider the first two symbols of a ternary data stream to be parsed and coded using the tree in Fig.~\ref{parsetree}.  If the first symbol is a $0$, the first index is used, that is, bits $000$ are encoded.  If the first symbol is not a $0$, the first two ternary symbols are represented as a single index --- $001$, $010$, etc.  So, for example, if Fig.~\ref{parsetree} is the (ternary) coding tree, then an input of $12100$ would first have $12$ parsed, coded into binary $011$; then have $10$ parsed, coded into binary $001$; then have $0$ parsed, coded into binary $000$, for an encoded output bitstream of $011001000$.

If ${X_k}$ is \textit{i.i.d.}, then the probability of $l_j$-symbol codeword $c(j) = c_1(j) \cdot c_2(j) \cdot c_3(j) \cdots c_{l_j}(j)$ is $\prod_{k=1}^{l_j} r_{c_k(j)}$ where
$r_j = \mbox{Pr}[X_{l_j} = j]$; i.e., 
internal nodes have probability equal to the product of their corresponding events.  An optimal tree will be that which maximizes expected compression ratio, the numbers of input bits divided by output bits.  The number of input symbols per parse is $l_j$ symbols ($(\log_2 D)l_j$ bits), depending on $j$, while the number of output bits will always be $\log_2 n$.  Thus the expected ratio to maximize is:
\begin{equation}
\sum_{j=0}^{n-1} r_j \frac{(\log_2 D)l_j}{\log_2 n} = \left(\log_n D\right) \sum_{j=0}^{n-1} r_j l_j
\label{maxval}
\end{equation}
where the constant to the left of the right summation term can be ignored, leaving expected input parse length as the value to maximize.

Because probabilities are fully known ahead of time, if we start with an optimal (or one-item) tree, the nature of any split of a leaf into other leaves, leading to a new tree with one more output item, is fully determined by the leaf we choose to split.  Since splitting increases expected length (the value to maximize) by the probability of the node split, we should split the most probable node at each step, starting with a null tree, until we get to an $n$-item tree.  Splitting one node does not affect the benefit value of splitting nodes that are not its descendents.  This greedy, inductive splitting approach is Tunstall's optimal algorithm.  Note that, because the output symbol is of fixed length, codeword probabilities should be as uniform as possible, which Tunstall's algorithm accomplishes via always splitting the most probable item into leaves, one of which will be the least probable item in the subsequent tree.

Tunstall's technique stops just before $n$ leaves are exceeded in the tree; this might have less than $n$ leaves as in Fig.~\ref{parsetree} for $n=8$.  The optimal algorithm necessarily has unused codes in such cases, due to the fixed-length nature of the output.  Markov sources can be parsed using a parameterized generalization of the approach where the parameter is determined from the Markov process, independent of code size, prior to building the tree\cite{Sava,TaRi}.

Analyses of the performance of Tunstall's technique are prevalent in the literature \cite{Abr01,TKR},
but perhaps the most obvious advantage to Tunstall codes is that of being randomly accessible\cite{TKR}: Each output block can be decoded without having to decode any prior block.  This not only aids in randomly accessing portions of the compression sequence, but also in synchronization: Because the size of output blocks is fixed, simple symbol errors do not propagate beyond the set of input symbols a given output block represents.  Huffman codes and variable-to-variable-length codes (e.g., those in \cite{BDS}) do not share this property.

Although much effort has been expended in the analysis of Tunstall codes and codec implementation, until recently few have analyzed the complexity of generating such codes.  The algorithm itself, in building a tree element by element, would be $O(n^2)$ time given a na\"{\i}ve implementation or $O(n \log n)$ time using a single priority queue.  Since binary output blocks are of size $\lceil \log_2 n \rceil$, this is somewhat limiting.  However, recently two independent works\cite{Kief,ReAn} showed that new algorithms based on that of Tunstall (and Khodak\cite{Khod}) could derive an optimal code in sublinear time (in the number of output items) given a Bernoulli (\textit{i.i.d.}~binary) input random variable.  

However, many input sources are not binary and many are not \textit{i.i.d.}; indeed, many are not even memoryless.  A more general linear-time algorithm would thus be of use.  Even in the binary case, these algorithms have certain drawbacks in the control of the construction of the optimal parsing tree.  As in Tunstall coding, this parsing tree grows in size, meaning that a sublinear algorithm must ``skip'' certain trees, and the resulting tree is optimal for some $n'$ which might not be the desired~$n$.  To grow the resulting tree to that of appropriate size, one can revert to Tunstall's tree-growing steps, meaning that they are --- and their implementation is --- still relevant in finding an optimal binary tree.


Here we present a realization of the original Tunstall algorithm that is linear time with respect to the number of output symbols.  This simple algorithm can be expanded to extend to nonidentically distributed and (suboptimally) to Markov sources.  Because such sources need multiple codes for different contexts, the time and space requirements for the algorithm are greater for such sources, although not prohibitive and still linear with the size of the output.  Specifically, if we have a source with $s$ states, then we need to build $s$ $D$-ary trees.  If the \textit{total} number of output leaves is $n$, then the algorithm presented here takes $O((1+(\log s)/D) n)$ time and $O(n)$ space.  (This reasonably assumes that $g \leq O(n)$, where $g$ is the number of possible triples of conditional probabilities, tree states, and node states; e.g., $g=2$ for Bernoulli sources and $g \leq Ds^2$ for any Markov input.)


\section{Linear-time Bernoulli algorithm} 
\label{memoryless} 

The method of implementing Tunstall's algorithm introduced here is somewhat similar to two-queue Huffman coding\cite{Leeu}, which is linear time given sorted probabilities.  The two-queue algorithm proceeds with the observation that nodes are \textit{merged} in \textit{ascending} order of their overall total probability.  Thus a queue can be used for these combined nodes which, together with a second queue for uncombined nodes, assures that the smallest remaining node can be dequeued in constant time from the head of one of these two queues.

In Tunstall coding, leaves are \textit{split} in \textit{descending} order.  Consider a node with probability $q$ split into two nodes: a left node of probability $a q$ and a right node of probability $(1-a) q$.  Because every prior split node had probability not exceeding $q$, the left child will have no larger a probability than any previously created left child, and the right child will have no larger a probability than any previously created right child.  Thus, given a Bernoulli input, it is sufficient to use two queues to have a linear-time algorithm for computing the optimal Tunstall code, as in Fig.~\ref{binaryfig}.

\begin{figure}
\textbf{~Linear-time binary Tunstall code generation}
\begin{enumerate}
\item Initialize two empty regular queues: $\overleftarrow{Q}$ for left children and $\overrightarrow{Q}$ for right children.  These queues will need to hold at most $n$ items altogether. 
\item Split the root (probability $1$) node with the left child going into $\overleftarrow{Q}$ and the right child going into~$\overrightarrow{Q}$.  Assign tree size $z \leftarrow 2$.
\item Move the item with the highest (overall) probability out of its queue.  The node this represents is split into its two children, and these children are enqueued into their respective queues.
\item Increment tree size by $1$, i.e., $z \leftarrow z+1$; if $z < n$, go to Step 3; otherwise, end.
\end{enumerate}
\caption{Steps for linear-time binary Tunstall code generation}
\label{binaryfig}
\end{figure}

\begin{figure*}
\psfrag{l}{{\tiny $0.7$}}
\psfrag{ll}{{\tiny $0.49$}}
\psfrag{lll}{{\tiny $0.343$}}
\psfrag{llr}{{\tiny $0.147$}}
\psfrag{lr}{{\tiny $0.21$}}
\psfrag{r}{{\tiny $0.3$}}
\begin{center}
\resizebox{11.5cm}{!}{\includegraphics{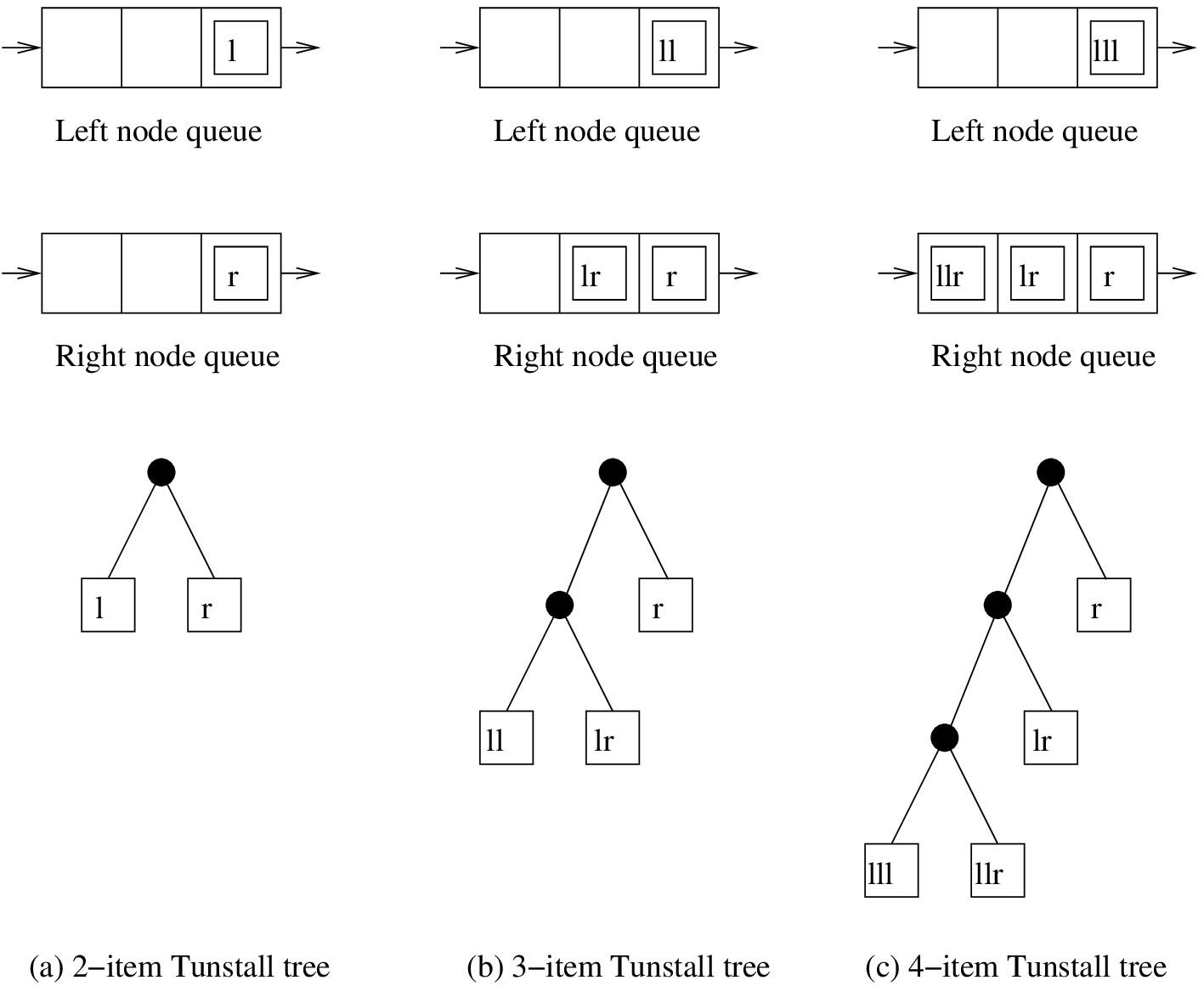}}
\caption{Example of binary Tunstall coding using two queues}
\label{tsplit}
\end{center}
\end{figure*}

\begin{example}
Consider the simple example of coding a Bernoulli$(0.7)$ input using a two-bit (four-leaf) tree, illustrated in Fig.~\ref{tsplit}.   Initially (Fig.~\ref{tsplit}a), the ``left'' queue has the left child of the root, of probability $0.7$, and the ``right'' queue has the right child, of probability~$0.3$.  Since $0.7$ is larger, the left node is taken out and split into two nodes: the $0.49$ node in the left queue and the $0.21$ node in the right queue (Fig.~\ref{tsplit}b).  The $0.49$ node follows (being larger than the $0.3$ node and thus all other nodes), leaving leaves of probability $0.343$ (last to be inserted into the left queue, corresponding to input $000$), $0.147$ (last in the right queue, input $001$), $0.21$ (input $01$) and $0.3$ (input $1$) (Fig.~\ref{tsplit}c).  The value maximized, compression ratio (\ref{maxval}), is 
$$(\log_4 2)\sum_{j=0}^3 r_i l_i = \frac{219}{200}=1.095.$$
\label{example1}
\end{example}
As with Huffman coding, allowing larger blocks of data generally improves performance, asymptotically achieving entropy; 
related properties are explored in \cite{DRSS}.

As previously indicated, there are faster methods to build optimal trees for Bernoulli sources.  However, these sublinear-time methods do not directly result in an optimal representation of a given size, instead resulting in one for a (perhaps different) output alphabet size not exceeding~$n$.  Any method that achieves a smaller optimal tree more quickly can therefore achieve an optimal $n$-leaf tree more quickly using the method introduced here in postprocessing.

\section{Fast generalized algorithm} 
\label{general}

This method of executing Tunstall's algorithm is structured in such a way that it easily generalizes to sources that are not binary, are not \textit{i.i.d.}, or are neither.  If a source is not \textit{i.i.d.}, however, there is state due to, for example, the nature or the quantity of prior input.  Thus each possible state needs its own parsing tree.  Since the size of the output set of trees is proportional to the total number of leaves, in this case $n$ denotes the total number of leaves.

In the case of sources with memory, a straightforward extension of Tunstall coding might not be optimal\cite{SaGa}.  Indeed, the optimal parsing for any given point should depend on its state, resulting in multiple parsing trees.  Instead of splitting the node with maximum probability, a \textit{generalized Tunstall policy} splits according to the node maximizing some (constant-time-computable) $f_{j,k}(p_i^{(j)})$ \textit{across all parsing trees}, where $j$ indexes the beginning state (the parse tree) and $k$ indexes the state corresponding to the node; $p_i^{(j)}$ is the probability of the $i$\textsuperscript{th} leaf of the $j$\textsuperscript{th} tree, conditional on this tree.  Every $f_{j,k}(p)$ is decreasing in $p = p_i^{(j)}$, the probability corresponding to the node in tree $j$ to be split.  This generalization generally gives suboptimal but useful codes; in the case of \textit{i.i.d.} sources, it achieves an optimal code using $f_{j,k}(p)=-\ln p$, where $\ln$ is the natural logarithm. 
The functions $f_{j,k}$ are yielded by preprocessing which we will not count as part of the algorithm time cost, being independent of $n$.  In this case $n$, the size of the output, is actually the number of total leaves in the output set of trees, not the number in any given tree.  These functions are chosen for coding that is, in some sense, asymptotically optimal\cite{TaRi}. 

Consider $D$-ary coding with $n$ outputs and $g$ equivalent output results in terms of states and probabilities --- e.g., $g=3$ for \textit{i.i.d.} input probability mass function $(0.5, 0.2, 0.2, 0.1)$, since events all have probability in the $g$-member set $\{0.5, 0.2, 0.1\}$.  If the source is memoryless, we always have $g \leq D$.  A more complex example might have $D$ different output values with different probabilities with $s$ input states and $s$ output states, leading to $g=Ds^2$.  Then a straightforward extension of the approach, using $g$ queues, would split the minimum-$f$ node among the nodes at the heads of the $g$ queues.  This would take $O(n)$ space and $O((s+g/D)n)$ time per tree, since there are $\lceil (n-1)/(D-1) \rceil$ steps with $g$ looks (for minimum-$f_{j,k}$ nodes, only one of which is dequeued) and $D$ enqueues as a result of the split of a node into $D$ children.  (Probabilities in each of the multiple parsing trees are conditioned on the state at the time the root is encountered.)

However, $g$ could be large, especially if $s$ is not small.  We can instead use an $O(\log g)$-time priority queue structure --- e.g., a heap --- to keep track of the leaves with the smallest values of $f$.  Such a priority queue contains up to $g$ pointers to queues; these pointers are reordered after each node split from smallest to largest according to priority $f_{j,k}(p_*^{(j)})$, the value of the function for the item at the head of the corresponding regular queue.  (Priority queue insertions occur anywhere within the queue that keeps items in the queue sorted by priorities set upon insertion.  Removal of the smallest $f$ and inserts of arbitrary $f$ generally take $O(\log g)$ amortized time in common implementations\cite[section~5.2.3]{Knu3}, although some have constant-time inserts\cite{FrTa}.)  The algorithm --- taking $O((1+(\log s)/D) n)$ time and $O(n)$ space per tree, as explained below --- is thus as described in Fig.~\ref{generalfig}.

\begin{figure}
\textbf{~Efficient coding method for generalized Tunstall policy}
\begin{enumerate}
\item Initialize empty regular queues $\{Q_t\}$ indexed by all $g$ possible combinations of conditional probability, tree state, and node state; denote a given triplet of these as~$t = (p', j, k)$.  These queues, which are not priority queues, will need to hold at most $n$ items (nodes) altogether. 
Initialize an additional empty priority queue $P$ which can hold up to $g$ pointers to these regular queues.
\item Split the $s$ root (probability $1$) nodes among regular queues according to~$(p', j, k)$.  Similarly, initialize the priority queue to point to those regular queues which are not empty, in an order according to the corresponding~$f_{j,k}$.  Assign solution size $z \leftarrow Ds$.
\item Move the item at the head of $Q_{P_0}$ --- the queue pointed to by the head $P_0$ of the priority queue~$P$ --- out of its queue; it has the lowest $f$ and is thus the node to split.  Its $D$ children are distributed to their respective queues according to~$t$.  Then $P_0$ is removed from the priority queue, and, if any of the aforementioned children were inserted into previously empty queues, pointers to these queues are inserted into the priority queue.  $P_0$, if $Q_{P_0}$ remains nonempty, is also reinserted into the priority queue according to $f$ for the item now at the head of its associated queue.

\item Increment solution size by $D-1$, i.e., $z \leftarrow z+D-1$.  If $z \leq n-D+1$, go to Step 3; otherwise, end.
\end{enumerate}
\caption{Steps for efficient coding using a generalized Tunstall policy} 
\label{generalfig}
\end{figure}

As with the binary method, this splits the most preferred node during each iteration of the loop, thus implementing the generalized Tunstall algorithm.  The number of splits is $\lceil (n-1)/(D-1) \rceil$ and each split takes $O(D + \log g)$ time amortized.  The $D$ factor comes from the $D$ insertions into (along with one removal from) regular queues, while the $\log g$ factor comes from one amortized priority queue insertion and one removal per split node.  While each split takes an item out of the priority queue, as in the example below, it does not necessarily return it to the priority queue in the same iteration.  Nevertheless, every priority queue insert must be one of either a pointer to a queue that had been previously removed from the priority queue (which we amortize to the removal step) or a pointer to a queue that had previously never been in the priority queue (which can be considered an initialization).  The latter steps --- the only ones that we have left unaccounted --- number no more than $g$, each taking no more than $\log g$ time, so, under the reasonable assumption that $g \log g \leq O(n)$, these initialization steps do not dominate. (If we use a priority queue implementation with constant amortized insert time, such as a Fibonacci heap\cite{FrTa}, this sufficient condition becomes $g \leq O(n)$.)

We thus have an $O((1+(\log g)/D) n)$-time method ($O((1+(\log s)/D) n)$ in terms of $n$, $s$, and $D$, since $g \leq Ds^2$) using only $O(n)$ space to store the tree and queue data.  The significant space users are the output trees ($O(n)$ space); the queues ($g$ queues which never have more items in them total than there are tree nodes, resulting in $O(n)$ space); and the priority queue ($O(g)$ space).

\begin{figure*}
\psfrag{1}{{\large $1$}}
\psfrag{2}{{\large $2$}}
\psfrag{Q15}{$Q_{(0.5,1,2)}$}
\psfrag{Q14}{$Q_{(0.4,1,2)}$}
\psfrag{Q13}{$Q_{(0.3,1,2)}$}
\psfrag{Q12}{$Q_{(0.25,1,1)}$}
\psfrag{Q25}{$Q_{(0.5,2,2)}$}
\psfrag{Q24}{$Q_{(0.4,2,2)}$}
\psfrag{Q23}{$Q_{(0.3,2,2)}$}
\psfrag{Q22}{$Q_{(0.25,2,1)}$}

\psfrag{f13}{{\scriptsize $f_{1,2}(0.3)$}}
\psfrag{f25}{{\scriptsize $f_{2,2}(0.25)$}}
\psfrag{f22}{{\scriptsize $f_{2,1}(0.25)$}}
\psfrag{f15}{{\scriptsize $f_{1,2}(0.2)$}}
\psfrag{f12}{{\scriptsize $f_{1,1}(0.1)$}}
\psfrag{g13}{{\scriptsize $=1.15\ldots$}}
\psfrag{g25}{{\scriptsize $=1.38\ldots$}}
\psfrag{g22}{{\scriptsize $=1.43\ldots$}}
\psfrag{g15}{{\scriptsize $=1.56\ldots$}}
\psfrag{g12}{{\scriptsize $=2.30\ldots$}}


\psfrag{0.5}{$0.5$}
\psfrag{\{0.25, 0.25\}}{$\{0.25,0.25\}$}
\psfrag{\{0.4, 0.3, 0.3\}}{$\{0.4,0.3,0.3\}$}

\psfrag{2l}{{\scriptsize $0.5$}}
\psfrag{1l}{{\scriptsize $0.4$}}
\psfrag{2ll}{{\scriptsize $0.25$}}
\psfrag{1ll}{{\scriptsize $0.2$}}
\psfrag{2lc}{{\scriptsize $0.125$}}
\psfrag{1lc}{{\scriptsize $0.1$}}

\psfrag{1r}{{\scriptsize $\overrightarrow{0.3}$}}
\psfrag{2r}{{\scriptsize $\overrightarrow{0.25}$}}
\psfrag{2lr}{{\scriptsize $\overrightarrow{0.125}$}}
\psfrag{1lr}{{\scriptsize $\overrightarrow{0.1}$}}

\psfrag{1c}{{\scriptsize $\underline{0.3}$}}
\psfrag{1cl}{{\scriptsize $\underline{0.15}$}}
\psfrag{1cc}{{\scriptsize $\underline{0.075}$}}
\psfrag{1cr}{{\scriptsize $\underline{\overrightarrow{0.075}}$}}
\psfrag{2c}{{\scriptsize $\underline{0.25}$}}

\psfrag{g}{{\large $g$}}
\begin{center}
\resizebox{16cm}{!}{\includegraphics{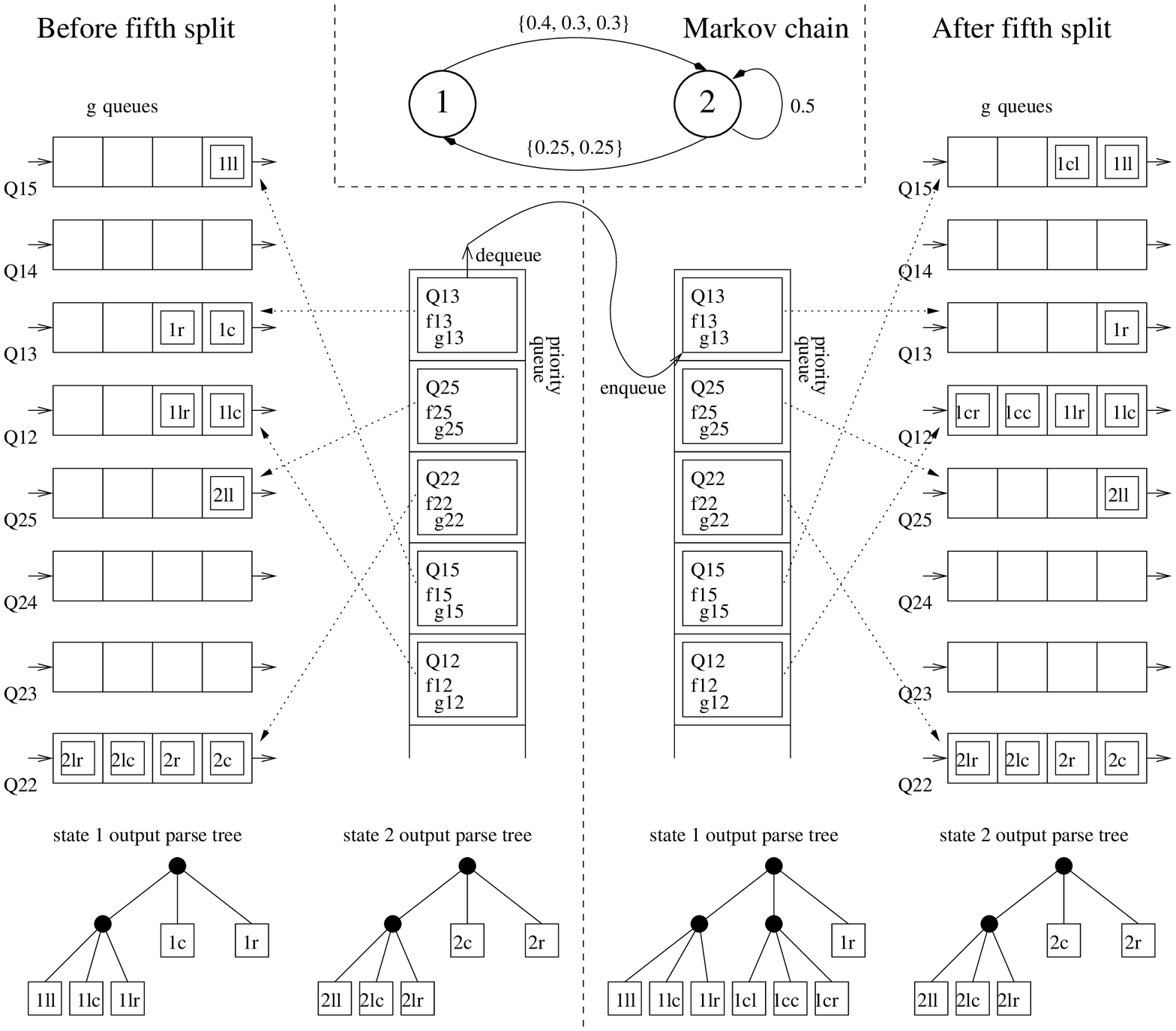}}
\caption{Example of efficient generalized Tunstall coding for Markov chain (top-center) shown before (left) and after (right) the fifth split node.  Right arrow overscore denotes right-most leaf and underscore denotes center subtree (to distinguish items); $f_{j,k}$ denotes priority function.}
\label{gensplit}
\end{center}
\end{figure*}

\begin{example}
Consider an example with three inputs --- $0$, $1$, and $2$ --- and two states --- $1$ and $2$, according to the Markov chain shown in Fig.~\ref{gensplit}.  State $1$ always goes to state $2$ with input symbols of probability $p_0^{(1)} = 0.4$, $p_1^{(1)} = 0.3$, and $p_2^{(1)} = 0.3$.  For state $2$, the most probable output, $p_0^{(2)} = 0.5$, results in no state change, while the others, $p_1^{(2)} = 0.25$ and $p_2^{(2)} = 0.25$, result in a change back to state~$1$.  Because there are $2$ trees and each of $2$ states has $2$ distinct output probability/transition pairs, we need $g = 2 \times 2 \times 2$ queues, as well as a priority queue that can point to that many queues. Let $f_{1,1}(p) = f_{2,2}(p) = -\ln p$, $f_{1,2}(p) = -0.0462158-\ln p$, and $f_{2,1}(p) = 0.0462158-\ln p$. 

The fifth split in using this method to build an optimal coding tree is illustrated by the change from the left-hand side to the right-hand side of Fig.~\ref{gensplit}.  The first two splitting steps split the two respective root nodes, the third splits the probability $0.5$ node, and the fourth splits the probability $0.4$ node.  At this point, the priority queue contains pointers to five queues. (The order of equiprobable sibling items with the same output state does not matter for optimality, but can affect the output; for the purposes of this example, they are inserted into each queue from left to right.)  In this example we denote these node queues by the conditional probability of the nodes and the tree the node is in. 
For example, the first queue, $Q_{(0.5,1,2)}$, is that associated with any node that is in the first tree and represents a transition from state $2$ to state $2$ (that of probability $0.5$).

Before the step under examination, the queue that is pointed to by the head of the priority queue is the first-tree queue of items with conditional probability $0.3$ (i.e., $Q_{P_0} = Q_{(0.3,1,2)}$) and tree probability $p=0.3$.  Thus the node to split is that at the head of this queue, which has lowest $f$ value $f_{1,2}(p) = 1.1578\ldots$. This item is removed from the priority queue, the head of the queue it points to is also dequeued, and the corresponding node in the first tree is given its three children.  These children are queued into the appropriate queue: For the most probable item --- probability $0.15$, conditional probability $0.5$ --- is queued into $Q_{(0.5,1,2)}$, while the two items both having probability $0.075$ and conditional probability $0.25$ are queued into~$Q_{(0.25,1,1)}$.  Finally, because the removed queue was not empty, it is reinserted into the priority queue according to the priority value of its head, still $f_{1,2}(0.3)=1.1578\ldots$. 
No other queue needs to be reinserted since none of the new nodes entered a queue that was empty before the step.  In this case, then, the priority queue is unchanged, and the queues and trees have the states given in right-hand side.
\end{example}


\ifx \cyr \undefined \let \cyr = \relax \fi

\end{document}